# UML Artifacts Reuse: State of the Art

Hamza Onoruoiza Salami*, Moataz A. Ahmed
Information and Computer Science Department,
King Fahd University of Petroleum and Minerals
Dhahran 31261, Saudi Arabia
e-mail: {hosalami, moataz}@kfupm.edu.sa

*Abstract*—The benefits that can be derived from reusing software include accelerated development, reduced cost, reduced risk and effective use of specialists. Reuse of software artifacts during the initial stages of software development increases reuse benefits, because it allows subsequent reuse of later stage artifacts derived from earlier artifacts. UML is the de facto modeling language used by software developers during the initial stages of software development such as requirements engineering, architectural and detailed design. This survey analyzes previous works on UML artifacts reuse. The analysis considers four perspectives: retrieval method, artifact support, tool support and experiments performed. As an outcome of the analysis, some suggestions for future work on UML artifacts reuse are also provided.

## I. INTRODUCTION

Software reuse is the creation of new software from existing software or software knowledge rather than building them from scratch [1]. It refers to developing software systems using components of previously developed software. These components need not be code fragments; they could be design, test data, part of a manual, or duration and cost estimates [2]. The benefits of reuse include accelerated development, reduced overall development cost, increased dependability, effective use of specialists, and reduced risk [3]. However, these benefits are not without any downside. The drawbacks of reuse include increased maintenance costs, 'not-invented-here' syndrome, lack of tool support, increased effort to create and maintain component libraries, and effort to find and adapt reusable components [2, 3].

Two types of reuse are possible: systematic or deliberate reuse, in which software components purposefully constructed for future reuse are utilized; and opportunistic or accidental reuse, in which developers realize that previously developed components can be used in new software products. Systematic reuse results in robust, well documented and thoroughly tested artifacts. However, it requires time, effort and resources which some organizations are unwilling to sacrifice because there are no guarantees that such components will be reused in the future [2]. Opportunistic reuse is simpler, but components may not be in the best form to be reused.

There are four phases of software reuse: representation, retrieval, adaptation and incorporation [4]. At the representation stage, a model of the new software component (query) is presented. During retrieval, a software component which is similar to the query, and whose adaptation cost is minimal is selected from the components library or repository. The retrieved component is modified to obtain a new component during adaptation. Finally, the new component is incorporated or integrated into the repository.

The Unified Modeling Language (UML) is a general-purpose modeling language for describing and designing software systems [5]. UML is maintained by the Object Management Group (OMG), a consortium of companies. OMG has defined two categories of UML diagrams: structure diagrams which document the static structure of system objects; and behavior diagrams which show the dynamic behavior (collaborations, activities and state histories) of system objects [6].

This research focuses on the reuse of UML artifacts for several reasons. 1) UML is the de facto modeling language for software developers. 2) Reuse of artifacts from initial stages of software life cycle is beneficial as it leads to subsequent reuse of later stage artifacts derived from earlier artifacts [7]. UML is widely used during the initial stages of software development such as requirements engineering, architectural and detailed design. 3) With the advent of Model Driven Engineering (MDE), models (including UML models) are considered as the main artifacts during software development [8]. The scope of this review is limited to discussing previous work on reusing UML models. A review of design pattern detection methods can be found in [9]. In addition, model versioning, comparison and clone detection are discussed in [8]. The rest of this paper is organized as follows: Section II discusses the retrieval techniques used for comparing query and repository artifacts. In Section III, the type of UML artifacts supported by each work is described. The reuse tool accompanying each work is the subject of Section IV. Section V analyzes the experiments presented in the reviewed studies. Finally, we conclude the paper in Section VI.

## II. RERTRIEVAL TECHNIQUES

This section categorizes UML artifacts retrieval methods by the techniques used to compare query and repository artifacts. These techniques include graph matching, use of ontology, Case Based Reasoning (CBR) techniques and Information Retrieval (IR) techniques.

*A. Graph Matching Techniques*

Since many UML artifacts are diagrams (e.g. class, sequence, state machine and activity diagrams), it is not surprising that several UML artifact retrieval approaches utilize graph matching/graph similarity. Graph matching measures the degree of similarity of the structural information contained in the artifacts that are compared. Usually, graph vertices represent artifact concepts (e.g. class, interface, object) while graph edges depict relations (e.g. UML relationships) between the concepts.

Robinson et al. [10, 11] applied graph matching in the retrieval of sequence diagrams. Both query and repository sequence diagrams were represented as conceptual graphs in which UML metamodel elements are encoded as vertices and UML metamodel associations are encoded as edges. A graph matching algorithm (SUBDUE [12]) was then applied to find the similarity between both conceptual graphs. Given a query sequence diagram, the algorithm finds similar substructures in repository sequence diagrams.

Park et al. [13] determined the similarity score of two sequence diagrams using an iterative graph matching algorithm. First, both diagrams are converted to Message-Object-Order-Graphs (MOOGs) in which there are nodes whenever messages are sent or received, and there are edges denoting message flow between objects and time flow within objects. The matching algorithm computes a similarity score between two MOOGs based on the idea that graph elements (nodes or edges) in two graphs are similar if their respective neighborhoods are similar.

In [14], the authors compute the structural similarity of two class diagrams using an inexact graph matching technique. Using a lookup table containing difference values for various types of class relationships, they compute the similarity score from the adjacency matrix representation of each class diagram. The adjacency matrix is obtained by considering classes as nodes, and relationships between them as edges.

Fig. 1 shows graph representations of UML artifacts. Computing graph similarity scores often involves identifying common portions of both graphs by examining all their subgraphs. In a graph with n nodes, there are $O(2^n)$ subgraphs [15]. Calculating graph similarity is computationally expensive. Not surprisingly, approaches that utilize graph matching usually avoid exhaustive comparisons. Robinson et al. [10, 11] allow users to supply two parameters for limiting the search for common substructures in sequence diagrams: beam, to restrict the search by breadth; and limit, to restrict the search by number of expansions. Authors in [13] and [14] both use iterative algorithms for computing graph similarity. The former approach requires the user to specify the number of iterations. In the later work the (re)user limits the search by supplying genetic algorithm (GA) parameters such as size of population in each generation and number of generations.

*B. Case Based Reasoning Techniques*

Case Based Reasoning (CBR) is a problem solving paradigm in which new problems are solved by reusing the solution to similar past problems [16]. In CBR, past experiences are stored as cases. Cases may be composed of five parts: a situation and its goal; a solution; the result of carrying out the solution; elucidation of results; and lessons learned [17]. The CBR cycle involves four main tasks: retrieval of one or more previous cases which is/are most similar to a new problem; reusing or adapting the information and knowledge contained in the retrieved case(s) to solve the new case; revising the proposed solution by evaluating it and repairing faults; and retaining the new experience by adding it to the knowledge base [17].

Gomes et al. [18-21] have used CBR and the WordNet lexical ontology to retrieve software designs. Past designs are stored as class diagrams (cases) in a repository (case base). Three types of objects can be retrieved: classes, interfaces and packages. Retrieval is carried out in two phases: a computationally inexpensive phase in which a fixed number of relevant cases are chosen by using WordNet relations to index the cases; and a computationally demanding ranking phase when previously chosen objects are ordered using similarity metrics. After retrieval, one or more retrieved cases is adapted to build a new case [22]. During the revision stage, the generated class diagram is verified to identify and remove any incoherent component [23]. Finally, during retention the system decides if the new case should be included in the case base. Redundant cases are not added to the case base because they increase the response time and storage requirements of the system [23].

In the ReDSeeDS project, artifacts from previously developed software are stored as cases comprising a problem (i.e., requirements), and a solution (i.e., architecture, design and implementation code) [24]. In essence, requirements act as case indexes that have links to corresponding artifacts [25]. Requirements are represented in a Requirements Specification Language (RSL) [26] in three possible formats: scenarios written in less formal natural language sentences; scenarios written in more formal constrained Structured English sentences; and using UML activity and sequence diagrams. During retrieval, requirements for a new project are compared with requirements of all projects in the case base. The most similar case is returned for reuse, based on the intuition that systems with similar requirements should have many artifacts in common. Computation of similarity scores for two requirements relies on the WordNet lexicon and either IR techniques (for natural language sentences) or graph matching techniques (for constrained language scenarios). During reuse (adaptation), a transformation engine generates new interaction diagrams, as well as application and business logic code for the new requirement, from the retrieved case [27].

*C. Ontology Based Tehniques*

Ontologies are ways of organizing information in order to promote sharing and reuse of knowledge derived from the information. They specify the set of concepts and the relationships among the concepts in a particular domain. The semantic relatedness of two concepts is often inversely proportional to the distance between them in the ontology. Using general-purpose lexical ontologies such as WordNet minimizes the time and effort required to collect and organize domain-specific ontologies. However, general ontologies do not incorporate technical or domain-specific knowledge. Consequently, a domain specific ontology was used in [28] to replace the WordNet ontology previously used in [18].

Two types of domain-specific ontologies have been used for class diagram retrieval in [29]. 'Application ontologies' were developed to measure semantic similarity between UML class diagram classifiers and relationships identifiers. A 'domain ontology' was then used to measure the semantic similarity between classifiers names. Overall similarity between class diagrams was computed as a weighted sum of both similarity scores. The 'application ontologies' are built once, while the 'domain ontology' is built for each new problem domain.

Many other studies have used ontologies/taxonomies to determine different semantic similarities. A taxonomy shows hierarchy relationships between concepts, whereas an ontology can have any type of relationship. A data type taxonomy (see Fig. 2) is used in [20] for computing the conceptual distance between data types. In [30], conceptual closeness (or similarity) between software artifacts was computed from different facets, each of which is represented by terms forming a taxonomy. In addition, WordNet- based similarity measures have been used for determining semantic similarity of class names [14, 18, 31], class method names [31], class attribute names [23, 31] and use case events [25, 32].

In [33] information contained in use case diagrams about actors and use cases, as well as additional information from users are integrated into an Ontology Web language (OWL) base ontology. The ontology is stored in a repository using a relational database system. In order to reuse past use case diagrams, the (re)user enters query parameters such as actor, use case and project names. The authors' tool queries the OWL ontology and returns a list of relevant use case diagrams.

*D. Information Retrieval Methods*

Information retrieval (IR) involves locating documents (usually text) satisfying an information need from large collection of documents [34]. Every day, millions of people apply IR techniques when they use web search engines. Generally, the data stored and manipulated by IR systems could be textual, audio, visual or multimedia documents. However IR techniques are applicable to software artifacts that contain considerable amount of text [24].

Authors in [32] computed the similarity between requirement specifications of query and repository models using IR techniques. The requirements were in the form of use case flow of events. Events occurring in a particular domain were grouped into clusters based on the lexical meaning of words describing them. Each use case's flow of events was represented using the vector space model (i.e. as a multi-dimensional vector). Each dimension represented the number of events belonging to a particular cluster. Finally, the cosine distance measure was used to determine the degree of similarity between the two requirement specifications.

In [35], IR techniques were used for scenario management and reuse. Each scenario was represented as a set of attributes: goals; authors; events; actors; actions; and episodes (i.e. named subsequence of events in a scenario).

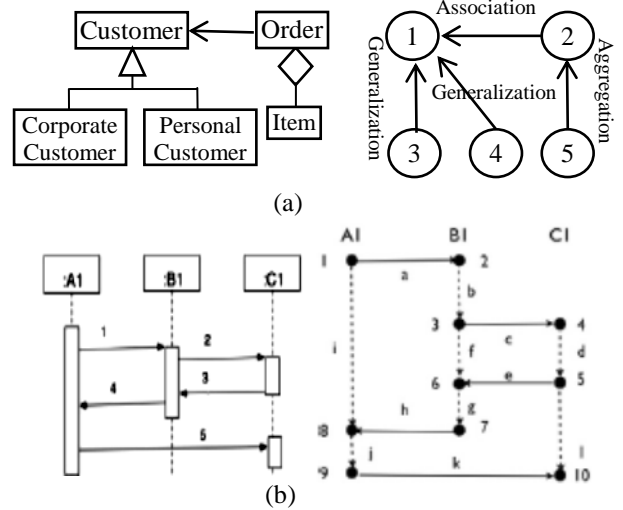

Figure 1. UML diagrams and their graphical representations (a) A Class Diagram and its corresponding directed graph [14]. (b) A Sequence Diagram and its corresponding Message-Object-Order-Graph (MOOG) [13].

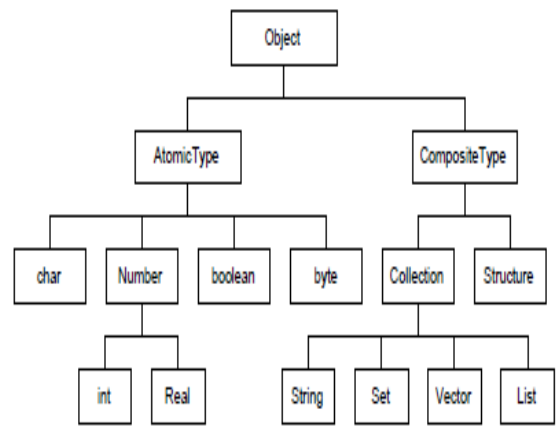

Figure 2. Data type Taxonomy [23]

The similarity between two scenarios was computed as the degree of overlap between their set of attributes using the Dice similarity coefficient.

Authors in [36] described a framework for retrieving UML artifacts from a repository in two stages: indexing and retrieval. During indexing a UML model in XMI form is stored in a relational database system. During retrieval, query and repository models are compared using either query inclusion or query similarity. The query inclusion technique searches for repository models that subsume the query model by formulating a SELECT statement and querying the database of repository models. On the other hand, query similarity is composed of topological and semantic similarity. Topological similarity is the Euclidean distance between two vectors where each vector dimension represents a different type of relationship and specifies the number of such relationships in a diagram. Semantic similarity is computed from the degree of overlap of terms occurring in both models and the distance (within a thesaurus) of terms/concepts occurring in relationships.

Ali and Du [30] have retrieved repository models in two steps; classification and retrieval. During classification a model is described from six perspectives/facets which capture the model's functional requirements. In the retrieval stage, a 'discrepancy ratio' similarity metric is computed using the degree of commonality in the descriptor terms of query and repository models.

*E. Other Methods*

This section describes other UML artifacts retrieval methods not discussed in previous sections. In [37], the similarity between sets of sequence diagrams is computed using two nested levels of genetic algorithm (GA). At the lower level, similarity is measured by mapping the classes in two sequence diagrams (using GA), and considering the number of matching and differing method calls. At the higher level, GA was used to map sequence diagrams in one model to sequence diagrams in the other model.

Park and Bae [13] determined the similarity score of two class diagrams using the Structure Mapping Engine (SME). The SME software works based on the structure mapping theory -an analogical mapping technique- which allows knowledge to be mapped in two domains by considering relational commonalities of objects in the domains regardless of the objects involved in the relationships.

Kotb [38] describes an approach for retrieving similar use case descriptions using textual entailment (TE), a natural language processing technique. A text T entails another text H if the meaning of H can be inferred from the meaning of T. With the aid of the WordNet lexicon, the author proposes comparing the summarized descriptions of query and repository use cases. Any repository use case whose summarized flow of events is entailed by that of the query is retrieved for reuse. During adaptation, a new scenario is generated from the entailed repository scenario using WordNet.

In [39] query and repository UML models are transformed from their XMI representations to specifications written in first order logic. The specifications are then matched, guided by some meta-knowledge (set of rules).

Authors in [40] have retrieved similar class diagrams from a repository using the class weights and the number of common class names in class diagrams. Each class is assigned a weight reflecting its degree of influence in the class diagram by considering the class's relationship types and the navigability and multiplicity of its association ends. Table I summarizes the retrieval techniques used in various studies. Each work is identified by its first author.

### III. ARTIFACT SUPPORT

UML diagrams are of two major kinds; structure diagrams and behavior diagrams [6]. Structure diagrams such as class, component, object, composite structure, deployment, package and profile diagrams document the static structure of system objects. On the other hand, behavior diagrams like activity, use case, state machine, sequence, communication, interaction overview and timing diagrams show the dynamic behavior of system objects. Of the fourteen different UML diagrams, each reuse work considers only a small number of diagrams. A summary of the UML diagrams considered for retrieval in each study is shown in Table II. Each work is identified by its first author.

The three UML diagrams that have been supported the most by various studies are class diagrams, sequence diagrams and use case diagrams. The flow of events in use case scenarios are almost always used rather than the use case diagrams themselves. Very little work has been done regarding the reuse of activity, state chart, collaboration and communication diagrams. Furthermore, nearly half of the UML diagrams have not been considered for reuse at all.

Only a few approaches consider more than one type of artifact during retrieval yet in most of these cases, general-purpose algorithms are used for multiple diagram types rather than specific algorithms that take into consideration the semantics of each UML diagram (see for example [10] and [30]). Park et al. [13]'s work stands out in this regard. The authors adopt a two-stage method for retrieving repository artifacts. In the first stage, class diagrams' structures are compared using analogy. Based on the structural similarity scores, a subset of repository UML models are selected. During the second stage, sequence diagrams in the shortlisted models are compared using a graph matching algorithm. The sequence diagrams are first converted to Message-Object-Order-Graphs in which edges represent message flow between objects and time flow within objects. The approach used in [13] is similar to the multi-view similarity method proposed in [41], where similarity of query and repository models is computed by considering structural, behavioral and functional views of UML models.

## IV. TOOL SUPPORT

In order to be effective, reuse techniques have to be complemented with software tools. These CASE tools can be either standalone programs or add-ins integrated to prominent CASE tools like Rational Rose and Enterprise Architect. Using the latter approach is more beneficial because large numbers of software developers use these CASE tools, and the tools maintain artifacts repositories which are essential for reuse. However, the standalone software reuse tools often accept UML models in XMI format, which is supported by many CASE tools. Table III shows the tools developed by various authors.

Most software reuse tools do not assist the reuser in adapting retrieved artifacts. In order to minimize the user's effort and save time, it is necessary to incorporate adaptation capability in reuse tools. Gomes et al. [18, 28]'s tools help users to create new class diagrams from the most similar existing diagram(s) [22]. Also, a transformation engine is used in the ReDSeeDS Engine to generate interaction diagrams, application logic and business logic code for new requirements, based on their similarity with existing repository requirements [27]. It is noteworthy that authors in [18, 27, 28] apply Case Based Reasoning (CBR) to software reuse. The reuse (adaptation) stage in CBR involves solving the new case using knowledge contained in retrieved cases.

TABLE I. SUMMARY OF RETRIEVAL TECHNIQUES USED IN VARIOUS STUDIES

| Work | Retrieval Techniques | | | | |
|---|---|---|---|---|---|
| | Graph matching | Ontology/ Taxonomy | Case Based Reasoning | Information Retrieval | Other Techniques |
| Ahmed 2006, [37] | | | | | Combinatorial optimization – Genetic Algorithm |
| Ali 2003, [30] | | √ | | √ | |
| Alspaugh 1999, [35] | | | | √ | |
| Bonilla-Morales 2012, [33] | | √ | | | Database Querying |
| Blok 1998, [32] | | √ | | √ | |
| Gomes 2002, [18] | | √ | √ | | |
| Khalifa 2008, [39] | | | | | Logic- First Order Logic |
| Kotb 2010, [38] | | √ | | | Natural language Processing- Textual Entailment |
| Llorens 2004, [36] | | √ | | √ | Database Querying |
| Park 2010, [13] | √ | | | | Analogy- Structure Mapping Theory |
| Robinson 2004, [10, 11] | √ | | | | |
| Robles 2012, [29] | | √ | | | |
| Rufai 2003 [31] | | √ | | | Combinatorial optimization – Hungarian Algorithm |
| Salami 2012, [14] | √ | √ | | | Combinatorial optimization – Genetic Algorithm |
| Wolter 2008, [24] ; Bildhauer 2009 [25] | √ | √ | √ | | |

TABLE II. UML ARTIFACTS SUPPORTED BY VARIOUS STUDIES

| Work | Class Diagram | Use Case Diagram | Sequence Diagram | Activity Diagram | Object Diagram | State Chart Diagram | Collaboration Diagram | Communication Diagram |
|---|---|---|---|---|---|---|---|---|
| Ahmed 2006, [37] | | | √ | | | | | |
| Ali Du 2003, [30] | √ | | √ | √ | √ | √ | √ | |
| Alspaugh 1999, [35] | | √ | | | | | | |
| Blok 1998, [32] | | √ | | | | | | |
| Bonilla-Morales 2012, [33] | | √ | | | | | | |
| Channarukul 2005, [40] | √ | | | | | | | |
| Gomes 2002, [18] | √ | | | | | | | |
| Khalifa 2008, [39] | √ | √ | √ | | | | | √ |
| Kotb 2010, [38] | | √ | | | | | | |
| Llorens 2004, [36] | √ | | | | | | | |
| Park 2010, [13] | √ | | √ | | | | | |
| Robinson 2004, [10, 11] | √ | √ | √ | | | | | |
| Robles et 2012, [29] | √ | | | | | | | |
| Rufai 2003, [31] | √ | | | | | | | |
| Salami 2012, [14] | √ | | | | | | | |
| Wolter 2008, [24] ; Bildhauer 2009 [25] | | √ | √ | √ | | | | |

TABLE III.    TOOL SUPPORT PROVIDED BY VARIOUS STUDIES

| Work | Tool | Type of Tool | | Adaptation of retrieved artifacts |
|---|---|---|---|---|
| | | *Stand-alone* | *Add-In* | |
| Bonilla-Morales 2012, [33] | | √ | | |
| Channarukul 2005, [40] | EASE DESIGN | √ | | |
| Gomes 2002, [18] | REBUILDER | √ | | √ |
| Gomes 2007, [28] | REBUILDER UML | | Enterprise Architect add-in | √ |
| Khalifa 2008, [39] | | √ | | |
| Robinson 2004, [10, 11] | SCENASST/REUSER | | Rational Rose add-in | |
| Robles 2012, [29] | | | Enterprise Architect add-in | |
| Smialek 2010, [27] | ReDSeeDS Engine | √ | | √ |

During retrieval, most reuse systems present ranked lists of a subset of repository models to the reuser. The models are listed in decreasing order of similarity to the query model. Reuse of the top model in the list should require the least amount of time and effort. Table IV presents the datasets and results obtained in different studies. The 'Results' column reports only precision and recall-based evaluation measures. For example, results in [31, 37] indicate that the similarity score obtained by comparing UML models of two versions of a software is high if the versions are close to each other, but low otherwise. Their results are not shown in Table IV.

Only few authors have evaluated their work using precision, recall or F measure. Reasonably high values of recall (100%) and precision (66-100%) are obtained in [13]. The F measure (71-94%) reported in [29] is impressive. Values of precision (21.4-80%) and recall (14.9-71.2%) obtained in [20] varies widely because 18 different experiments were performed using three versions of their retrieval algorithm and six different sizes of the retrieval set. The results in [20] clearly showed the inverse relationship existing between precision and recall values. The main difficulty involved in comparing results from different studies is that each study uses different query and repository models to evaluate their methods, due to the unavailability of openly accessible datasets from industry.

## V. CONCLUSION

In this review, we have discussed previous works on UML artifacts reuse from four perspectives: retrieval method, artifact support, tool support and experiments performed. Most studies combine more than one retrieval technique. The WordNet lexical ontology is commonly used in different studies because it is freely available thus minimizing knowledge gathering effort. However, it does not incorporate technical or domain specific knowledge. Case based reasoning holds promise for software reuse because the reuse (adaptation) phase allows automatic adaption of retrieved artifacts, thereby minimizing reuse time and effort.
Of the fourteen UML diagrams, only class diagrams, sequence diagrams and use case diagrams have been considered during retrieval by majority of the authors. Little or no work has been done regarding reusing the remaining UML diagrams.

## I. EXPERIMENTS

Empirical results obtained from experiments provide a good way to evaluate various UML reuse systems. In information retrieval, two measures commonly used to determine the effectiveness of unranked retrieval methods are precision and recall [34]. Both of these measures are used by several authors to present the results of their software reuse experiments. Precision is the proportion of retrieved artifacts/models that are relevant to the query. Recall is the proportion of relevant repository models that are retrieved. These two quantities tradeoff against each other. For example, high recall but low precision can be obtained by retrieving all repository models. On the other hand, high precision but low recall can be obtained by not retrieving any repository models [34, 42]. The F-measure is the weighted harmonic mean of precision and recall. It combines both measures into one value, depending on their relative importance.

Thus, it is imperative to direct some research effort to the reuse of other UML diagrams like activity diagrams and state machine diagrams.

The merits of multi-view measures which combine similarity measures of UML artifacts from different viewpoints (such as structural, behavioral and functional views) need to be investigated since software is typically modeled from multiple viewpoints [41]. In addition, retrieval techniques for the various UML diagrams should take into account the semantics of each different diagram, rather than make use of generic algorithms for multiple UML diagrams.

One of the factors limiting software reuse is lack of tool support. User friendly CASE tools need to accompany UML reuse techniques. Such tools should preferably be developed as add-ins to commonly used CASE tools like Enterprise Architect and Rational Rose which are used by large number of software designers/ developers.

Only few of the works discussed have evaluated their techniques using standard IR evaluation measures of precision, recall and F measure. Even in these few cases, the authors have used evaluation measures meant for unranked (set) retrieval rather than those specifically designed for ranked retrieval. Standard IR measures for evaluating ranked retrieval include *Mean Average Precision*, *R Precision* and 11-point average precision [34, 43]. In addition, the results presented in various studies cannot be compared because the studies use query and repository models of varying sizes and from different domains. Of greater concern is the fact that there is no evidence in the literature that any of these reuse systems has been/is being used in the industry.

ACKNOWLEDGMENT

The authors would like to acknowledge the support provided by the Deanship of Scientific Research at King Fahd University of Petroleum & Minerals (KFUPM) under Research Grant 11-INF1633-04.

TABLE IV.    SUMMARY OF EXPERIMENTS CARRIED OUT IN VARIOUS STUDIES

| Work | Query Artifact(s) | Repository Artifacts | Results |
|---|---|---|---|
| Ahmed 2006, [37] | Reverse engineered class diagrams for different versions of JDK and Apache Tomcat | Reverse engineered class diagrams for different versions of JDK and Apache Tomcat | |
| Ali 2003, [30] | Two models. One models for inventory control and another for online banking | One model for inventory tracking system | |
| Channarukul 2005, [40] | One class diagram | 12 class diagrams | |
| Blok 1998, [32] | One use case scenario from banking domain | 20 use case scenarios from banking domain | |
| Gomes 2003, [20] | 25 class diagrams | Sixty cases (class diagrams) from four domains: education, banking, health and stores (video stores and groceries) | Recall = 14.9 – 71.2%<br>Precision = 21.4 – 80% |
| Park 2010, [13] | 10 models from 'launcher-and-patcher'[a] domain | 10 models from 'launcher-and-patcher' domain | Recall = 100%<br>Precision = 66 – 100% |
| Robinson 2004, [10, 11] | Seven use cases and 27 classes from sales processing domain | 85 use cases and 308 classes from order processing domain | |
| Robles 2012, [29] | 15 class diagrams from education domain | 65 class diagrams from education domain | F measure = 71-94% |
| Rufai 2003 [31] | Reverse engineered class diagrams for different versions of JDK, J2SDK and Apache ANT | Reverse engineered class diagrams for different versions of JDK, J2SDK and Apache ANT | |

a. Launcher and patchers are software package management systems

* Corresponding Author:
Hamza Onoruoiza Salami,
Information and Computer Science Department,
King Fahd University of Petroleum and Minerals, Dhahran, Saudi Arabia,
Email: hosalami@kfupm.edu.sa    Tel:+966-3860-7356